# Towards extremely high-resolution broad-band flat-field  spectrometer in "water window"

Zhuo Li,[1,2] Bin Li,[1,2,3*]

[1]*Shanghai Institute of Applied Physics, Chinese Academy of Sciences, Shanghai 201204, China*

[2]*University of Chinese Academy of Sciences, Beijing 100049, China*

[3]*School of Physical Science and Technology, ShanghaiTech University, Shanghai 201210, China*

*Correspondence e-mail: libin1995@sinap.ac.cn*

**The optical design of a novel spectrometer is present, combining a cylindrically convex pre-mirror with a cylindrically concave VLS grating (both in meridional) to deliver a resolving power of 100,000-200,000 in 'water window' (2-5nm). More remarkably, unlike a typical RIXS spectrometer to obtain such high resolution by tight focusing or tiny confinement slit (<1μm), here the resolution could be achieved for an effective meridional source size of 50μm (r.m.s.). The overall optical aberrations of the system are well analysed and compensated, providing an excellent flat field at the detector domain throughout the whole spectral range. And a machine learning scheme - SVM is introduced to explore and reconstruct the optimal system with pretty high efficiency.**

## 1.Introduction

In the past few decades, X-ray spectrometers accomplished rapid development driven by advanced light sources such as synchrotron radiation facilities or free electron lasers, and were widely used for exploring various intriguing research topics especially in the regions of extreme ultraviolet or soft X-ray. A high-resolution spectrometer can help researchers to investigate the energy、momentum and polarization dependence of the photon-matter interaction or scattering processes, and hence to reflect the intrinsic properties of charge、spin、orbital、lattice excitation and etc[1]. With the further improvement of resolving power, instance charge transfer and d-d excitations[2, 3], spin excitations in cuprates[4, 5] and iron pnictides[6], high energy phonons[4], and vibrations in single molecules[7, 8] could be thoroughly investigated, and previously unobserved features and details of the materials' spectroscopy could also be revealed.

The 'water window', spanning the wavelength range of 2–5 nm, is able to provide the excellent contrast imaging for C or O atoms and related structures; this outstanding property could be utilized to image and analyze the biological cells or microstructures in vitro and potentially in vivo. 'Water window' spectroscopy is also a novel probe for material properties and electron energy states.

Previously, the high-resolution spectrometers in this spectral range include the following designs of role models: grating on Rowland circle structure[9]; single plane grating grooved in varied line spacing (VLS) [10, 11]; single concave VLS grating[12, 13]; a concave mirror pre-focusing the incident beam upstream a plane grating, creating a real secondary source[14]; the pre-focusing spherical mirror converges beam beyond the VLS grating, creating a virtual source, i.e. Hettrick Underwood design[15], which derived into different versions: e.g. i) C.F. Hague and J. H. Underwood employed a KB mirror for pre-focusing to correct spectral astigmatism[16]; ii) G. Tondello replaces the KB mirror to a toroidal mirror[17]; iii) Joseph Dvorak added a deflection mirror downstream the grating to level the outgoing beam[18]; iv) Hetrick Underwood scheme implementing a Wolter-type focusing system[19], and etc; Beside these, Yi-De Chuang and Yu-Cheng Shao designed a modular spectrometer whose modules could be conveniently adapted to various research requirements[20].

In the past, convex mirrors were rarely used, only Wolter III focusing system consisting of a hyperbolically convex mirror and an elliptically concave mirror is adopted in X-ray imaging and microscopy [21]. Where the incoming beam is grazing incident on the convex mirror and the reflection beam is diverging; its reverse extension lines are converged at one focus of the elliptical mirror, the reflected beam from the ellipse is propagating backward and then focused on the other focus. While except for a few reports[22, 23], the characteristics of the Wolter Type III mirrors are rarely studied, lack of deep and clear understanding. Inspired by Wolter configuration and based on these previous works, we formulated a delicate high-resolution flat-field spectrometer design in "water window", combining of an upstream pre-divergent convex mirror and a downstream concave VLS grating, which is demonstrated to enhance the resolving power considerably while maintain the decent flat field condition throughout the spectral range.

And we are aware of that a resonant inelastic X-ray scattering (RIXS) spectrometer usually owns a very high resolving power, benefiting from an excellent upstream monochromator system, via confining or focusing the beam through an exist slit of width down to ~1 μm, representing a pretty small secondary light source for the RIXS spectrometer[18]. So the super-high resolving power is achieved at the cost of a large amount beam flux while inducing significant optical aberration. While our efforts are completely different, aiming to achieve such high resolving power but without sacrificing the beam flux to the spectrometer, i.e. maintaining the original beam with a rather large source size. Then the intrinsic optical nature of the system, and the primary factors influencing the spectral distribution quality and resolution are explicitly analysed to exploit its best performance. And the manuscript is organized as follows:

a) The 2nd section presented the numerical simulation and algorithm to prove that the convex pre-mirror is a better choice for enhancing the resolving power of a spectrometer. Beside the resolution enhancement, the decent flat-field could be achieved at the detector, since the optical aberration of the convex mirror propagates downstream to compensate that of the concave grating, thus optimizing the primary aberration of the overall system.

b) The 3rd section explicitly discussed about the optimization algorithm, where the machine learning tool Support Vector Machine (SVM) is introduced and implemented to achieve a set of optimal parameters in the spectrometer design, while the quality evaluation parameter for the spectral imaging is well defined and discussed.

c) The 4th section mainly discusses about the key parameters of the system (e.g. source size, optical aberrations, fabrication errors etc.) determining the ultimate spectral resolution, which is verified by the ray-tracing program. Especially the critical requirement for the slope errors of the optical elements in the high-resolution spectrometer is analysed.

d) Finally, we made a more general and summarizing remark regarding to our design and discussed about the potential research and development in the future.

## 2. Numerical Simulation

Firstly, we listed a set of parameters fixed in the simulation and discussion throughout this article: i) light source intensity distribution: Gaussian profile; ii) size of the light source: $\sigma_S$ =50μm(r.m.s); iii) divergence angle of light source: 20μrad(r.m.s); iv) grooved density of the VLS grating at the center $D_0$ =24000ln/cm; v) grating diffraction order, $m$=1; vi) wavelength range: 2~5nm (water window); vii) the distance from the original light source to the grating $L$=30m=3000cm, etc. Here we are mainly concerned of the beam properties in its meridional coordinate, thus cylindrical substrates (tangentially convex or concave profiles) are adopted for all the optical elements in the system. This is sensible, since the beam divergences of synchrotron radiation or free electron laser are quite small, a free propagation beam in sagittal coordinate wouldn't lead to a large foot-print in that direction.

### 2.1 Four types of spectrometer

The single concave VLS grating spectrometer is shown in Fig.1(a), and its ideal resolving power is given by[24]:

$$A_1 = \frac{\lambda r D_0}{\sigma_s^{FWHM} \cos\alpha} = \frac{\lambda D_0}{\sigma_s^{FWHM} \cos\alpha} \times L \quad (1)$$

Where $\lambda$ is the wavelength, $r$ is the object distance of grating, $D_0$ (grating groove density) is defined previously, $\sigma_s^{FWHM}$ is the original source size in FWHM ($\sigma_s^{FWHM} \approx 2.355\sigma_s^{rms}$), and $\alpha$ is the incident angle of grating. Since $L$ is the distance from the original light source to the grating, so $r=L$ for this case (denoted by the dotted line arrow). According to Eq. (1), the resolving power is proportional to the wavelength, the groove density of the grating, the grating object distance $r$ (or $L$), inversely proportional to the size of light source size, and prefers a larger incident angle (or a smaller grazing incidence angle in complementary).

As shown in Fig.1(b), the concave VLS grating is combined with a pre-focusing concave mirror, forming a real secondary source for the grating i.e. the meridional beam focuses upstream the grating and illuminates it. So, the resolving power is calculated by:

$$A_2 = \frac{\lambda \cdot D_0 (d - r_c')}{(\sigma_s^{FWHM} M_c)\cos\alpha}$$
$$= \frac{\lambda D_0 \cdot (d - (L-d)M_c)}{(\sigma_s^{FWHM} M_c)\cos\alpha}$$
$$= \frac{\lambda D_0}{\sigma_s^{FWHM} \cos\alpha}(d(1+\frac{1}{M_c}) - L) \quad (2)$$

Where $r_c$ and $r_c'$ are the object and image distances of the pre-focusing mirror, whose magnification is denoted by $M_c = r_c'/r_c > 0$ (since $r_c' > 0$ for this case), $d$ is the separation in-between the concave mirror and the grating. So the object distance of the concave mirror is $r_c = L-d$, the grating object distance can be expressed as $r = d - r_c' > 0$, and the effective source size of the grating is $\sigma_s^{FWHM} M_c$.

Fig1.(c) presents a similar configuration as Fig.1(b), while the pre-concave mirror forms a virtual source for the grating, i.e. the meridional beam focuses behind the grating. This recalls the typical Hetrick-Underwood scheme, associate with a resolving power of:

$$A_3 = \frac{\lambda \cdot D_0 (-1)(d - r_c')}{(\sigma_s^{FWHM} M_c)\cos\alpha} = \frac{\lambda D_0}{\sigma_s^{FWHM} \cos\alpha}(L - d(1+\frac{1}{M_c})) \quad (3)$$

Where $M_c > 0$ (since $r_c' > d > 0$), the (-1) term in the numerator indicates the virtual source for the grating, and its object distance is $r = d - r_c' < 0$ (virtual source). The rest of variables in Eq. (3) are defined in a similar way as in Eq. (2).

Finally in Fig1.(d), the VLS grating is combined with a pre-convex mirror. The incident beam is diverged meridionally by the cylindrical convex mirror, and the virtual image of the convex mirror represents the real source of the grating effectively. The resolving power of the system is:

$$A_4 = \frac{\lambda D_0 \cdot (d - r_c')}{((-1)\sigma_s^{FWHM} M_c)\cos\alpha} = \frac{\lambda D_0}{\sigma_s^{FWHM} \cos\alpha}(L - d(1+\frac{1}{M_c})) \quad (4)$$

Where $M_c < 0$ since the pre-convex mirror generates a virtual image (the image distance $r_c' < 0$), and the object distance of the grating $r = d - r_c' > d > 0$. Similarly, the (-1) term in denominator of Eq. (4) is due to the virtual image of the convex mirror.

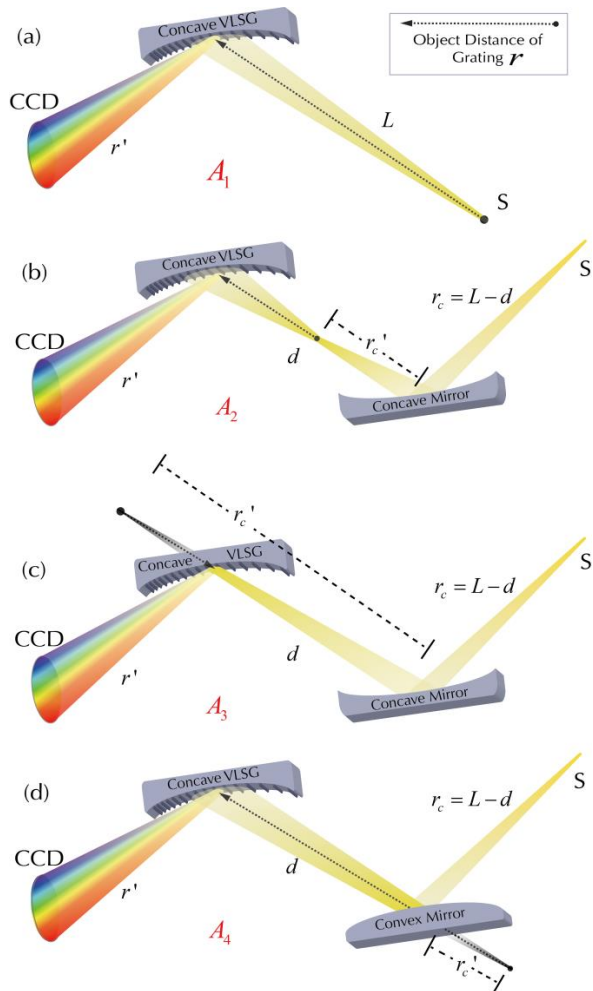

Fig. 1 Scheme diagram of four kinds of spectrometer designs implementing concave VLS grating with or without pre-mirror (corresponding to $A_1 \sim A_4$). S represents the light source, $L$ is the distance from the original light source to the grating, $d$ is the distance between the pre- concave (or convex) mirror and the grating, $r_c$ and $r_c'$ are the object and image distances of the concave (or convex) mirror respectively. $r$ is the object distance of grating indicated by a dotted arrow, $r'$ is the image distance of grating. (a) Single concave VLS grating, where the object distance of grating is $r=L$; (b) The concave VLS grating is combined with a pre-focusing concave mirror, forming a real source for the grating, $r = d - r_c' > 0$; (c) The similar case to (b), where the pre- concave mirror forms a virtual source for the grating, $r = d - r_c' < 0$; (d) The concave VLS grating is combined with a pre-diverged convex mirror, where the source of the grating is real, i.e. $r = d - r_c' > d > 0$ since $r_c' < 0$.

In order to evaluate the performance of these four systems, their resolving powers (refer to Eq. (1-4)) are plot against $M_c$ in Fig.2, for a set of 3 different optical element spacings, $d$=600, 1000, 1400cm. Again, the pre-set parameters at the beginning of the section 2 are used for the calculation, e.g. $L$=3000cm, $\sigma_s$=50μm (r.m.s), $\alpha$=89°, $D_0$=24000ln/cm. Since the resolving power is wavelength dependant, here only the results for $\lambda$=5nm are presented.

In Fig.2, $A_1$ (blue) is the control group with a constant resolution of ~170,000, where $M_c$ is not applicable since only a single concave grating is used in the system. And for the other three configurations, only if the value of $A_2 \sim A_4$ are greater than $A_1$, the resolving power could be regard as "enhanced". For $A_2$ (three green curves crossing the center of the graph vertically, with only minor differences in color), the resolving powers monotonously decrease with $M_c$ for each $d$, only if $M_c$ is less than 0.304, the resolution would be greater than $A_1$ (for $d$=1400cm; and $M_c$<0.200 for $d$=1000cm; $M_c$<0.111 for $d$=600cm). On the other hand, with $M_c$ increases, the focus of the pre-focusing mirror will gradually move to the surface of the grating, at that circumstance the resolving power declines down to the zero. Further increasing $M_c$, the system will transit to $A_3$, i.e. Hetrick Underwood design (yellow curves in bottom-right corner), where the focal spot behind the grating is corresponding to a virtual source of the grating. $A_3$ monotonously increases with $M_c$ for all $d$ values, and apparently smaller $d$ is associated with a relatively higher resolving power. However, since $A_3$ is always less than $A_1$ for any case, $A_3$ is unable to enhance the resolving power. For $A_4$ (three red curves in top-left corner), the pre-convex mirror generates a virtual image, i. e. $r_c' < 0$ and $M_c < 0$. And it is observed that $A_4$ monotonously decreases with $|M_c|$ for all $d$ values. When $|M_c| < 1$, the resolving power would be enhanced ($A_4 > A_1$). Especially when $|M_c|$ becomes smaller than 0.3 (the region confined by vertical dashed lines), $A_4$ would gain significant increasing (similar as $A_2$). But it needs to point out that too-much small value of $|M_c|$, generally associated with unacceptably large optical aberration delivered by the pre- focusing ($A_2$) or diverging ($A_4$) mirrors, should be avoided in the system design. According to Fig. 2 and the discussion above, $A_2$ can only achieve resolving power enhancement within the region of $|M_c| < 0.3$, while $A_4$ could achieve this outside the region, having a broader flexibility for system design. Therefore, configuration $A_4$ with a pre-convex mirror was chosen to develop an optimal spectrometer with enhanced resolving power (respected to $A_1$).

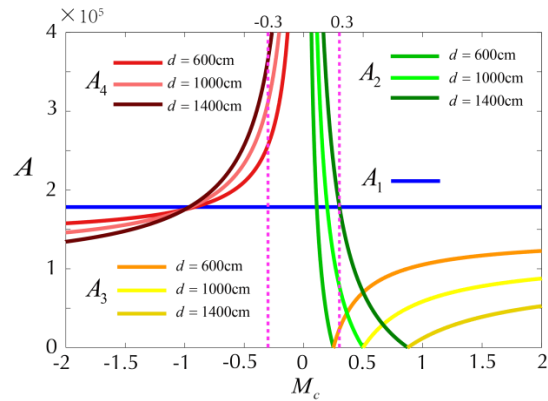

Fig. 2 Demonstrates the various resolving power $A_1 \sim A_4$ dependence on the magnification $M_c$ provided by the pre-mirror, at 3 typical optical elements spacing $d$ (separation in-between the pre-mirror and the grating), each of $A_1 \sim A_4$ are grouped into close colours. The common parameters for each case are: $L$=3000cm, $\lambda$=5nm, $\sigma_s$=50μm (r.m.s), $\alpha$=89°, $D_0$=24000ln/cm. $A_1$ (blue) is a constant about 170000; $A_2$ (green) would enhance the resolution only when $M_c$<0.3 (with respected to $A_1$); $A_3$ (yellow) cannot improve the resolution for all possible values of $M_c$; $A_4$ (red) can enhance the resolution as long as $|M_c| < 1$, where the region of $-0.3 < M_c < 0$ (or $0 < M_c < 0.3$ for

$A_2$) within the two purple dashed line is corresponding to extremely high resolution while the optical aberration for the pre- diverging (focusing) mirror is unacceptably high. $A_4$ is a preferable solution to enhance the resolving power when $|M_c|<1$ but not too much small.

### 2.2 Resolution enhanced flat-field spectrometer

We need to proceed the following steps to design an enhanced flat-field spectrometer, using configuration $A_4$.

a) Establish a set of fundamental parameters (refer to the beginning of section 2): Gaussian-light source; source size $\sigma_s^{rms}$ =50μm (r.m.s); source divergence angle: 20μrad (r.m.s); $D_0$ =24000ln/cm; $m=1$; wavelength range: 2~5nm; $L=3000$cm; and optical elements with meridionally cylindrical profiles.

b) Determine the image distance of grating $r'$

The magnification of a diffraction grating is:

$$M_g = \frac{\sigma_{CCD}^{(FWHM)}}{M_c \sigma_s^{(FWHM)}} = \frac{r'}{r}\frac{\cos\alpha}{\cos\beta} \quad (5)$$

Where the minimum value of $\sigma_{CCD}^{(FWHM)}$ is set to the pixel size of the CCD, which is the spatial limit to resolve the spectral distribution at the detector; $M_c\sigma_s^{(FWHM)}$ represents the effective source size of the grating created by the pre-convex mirror; $\alpha$ and $\beta$ are the incidence and diffraction angles of the grating respectively.

From the previous discussion, the object distance of grating is $r=(d-(L-d)M_c)$ (where $-1<M_c<0$; case $A_4$ in Fig.2). Then the image distance of grating should meet the following requirement:

$$r' \geq \frac{\sigma_{CCD}^{(FWHM)}}{\sigma_s^{(FWHM)}}\frac{\cos\beta}{\cos\alpha}[d(1+\frac{1}{M_C})-L] \quad (6)$$

So $r'$ is a function of $d$ and $M_c$, and could be interpreted as: an upstream pre-convex mirror creates a new light source and a new effective object distance for the grating, which determines the minimal image distance the grating should have.

c) Achieve the 'flat field'

The groove density of a VLS grating is:

$$n(w) = D_0 + D_1 w + D_2 w^2 + D_3 w^3 \quad (7)$$

Where the VLS coefficients $D_i$ could be optimized through the elimination of optical aberrations in various orders for the system, using the scheme we developed previously [24]. In addition, the grating on a cylindrically concave substrate with optimized VLS coefficients allows the achievement of an excellent meridional 'flat field' at its detector plane.

According to Fermat's principle for geometrical optics, the optimal imaging in meridional coordinates could be achieved through zeroing the first-order derivative of the light-path function connecting the light source and the image via optics (since the grating is a dispersive optic, various wavelengths are associated with different preferable optical paths) [25]. Especially the $F$ terms e.g. the first few dominants, should satisfy the following equations crossing the wavelength range:

$$F_{100} = -\sin\alpha - \sin\beta + D_0 m\lambda \quad (8)$$

$$F_{200} = \frac{1}{2}\left(\frac{\cos^2\alpha}{r}-\frac{\cos\alpha}{R}\right)+\frac{1}{2}\left(\frac{\cos^2\beta}{r'}-\frac{\cos\beta}{R}\right)-D_1 m\lambda\frac{1}{2} \quad (9)$$

$$F_{300} = \left(\frac{\cos^2\alpha}{r}-\frac{\cos\alpha}{R}\right)\frac{\sin\alpha}{2r}+\left(\frac{\cos^2\beta}{r'}-\frac{\cos\beta}{R}\right)\frac{\sin\beta}{2r'}-D_2 m\lambda\frac{1}{3} \quad (10)$$

$$F_{400} = \frac{1}{8}\left[\frac{4\sin^2\alpha}{r^2}\left(\frac{\cos^2\alpha}{r}-\frac{\cos\alpha}{R}\right)-\frac{1}{r}\left(\frac{\cos^2\alpha}{r}-\frac{\cos\alpha}{R}\right)^2+\frac{1}{R^2}\left(\frac{1}{r}-\frac{\cos\alpha}{R}\right)\right.$$
$$\left.+\frac{4\sin^2\beta}{r'^2}\left(\frac{\cos^2\beta}{r'}-\frac{\cos\beta}{R}\right)-\frac{1}{r'}\left(\frac{\cos^2\beta}{r'}-\frac{\cos\beta}{R}\right)^2+\frac{1}{R^2}\left(\frac{1}{r'}-\frac{\cos\beta}{R}\right)\right.$$
$$\left.-2D_3 m\lambda\right] \quad (11)$$

Where $R$ is the cylindrical radius of the grating. More specifically, the equation of $F_{100}$ is actually the grating formula; $F_{200}$ is related to the meridional focus, and could be utilized to characterize the 'defocus' over the whole spectral range; and $F_{300}$ and $F_{400}$ are associated with the 'coma' and 'spherical aberration', respectively.

The imaging distance of the grating which achieves the optimal flat field for the entire spectral range, according to[24]:

$$r'(\lambda) = \frac{\cos^2\beta(\lambda)}{D_1 m\lambda - \left(\frac{\cos^2\alpha}{r}-\frac{\cos\alpha}{R}\right)+\frac{\cos\beta(\lambda)}{R}} \quad (12)$$

Each set of the parameters would lead to a unique optimal meridional radius $R$ and coefficient $D_1$ only, then $D_2$ and $D_3$ could be derived at the central wavelength $\lambda_0$ by letting $F_{300}(\lambda_0)=0$ and $F_{400}(\lambda_0)=0$ via Eq. (10-11).

d) Correction of aberrations.

The above discussion is only applicable to a single concave grating. In case a pre-focusing (divergent) mirror is implemented in the system, the optical aberrations propagation from the upstream mirror need to take into account.

The primary aberrations of an upstream convex mirror could be calculated in a similar way as Eq. (9-11), using the optical path function and the relevant $F$-terms:

$$F_{200\_c} = \frac{1}{2}\left(\frac{\cos^2\alpha_c}{r_c}-\frac{\cos\alpha_c}{R_c}\right)+\frac{1}{2}\left(\frac{\cos^2\alpha_c}{r_c'}-\frac{\cos\alpha_c}{R_c}\right) \quad (13)$$

$$F_{300\_c} = \left(\frac{\cos^2\alpha_c}{r_c}-\frac{\cos\alpha_c}{R_c}\right)\frac{\sin\alpha_c}{2r_c}+\left(\frac{\cos^2\alpha_c}{r_c'}-\frac{\cos\alpha_c}{R_c}\right)\frac{\sin(-\alpha_c)}{2r_c'} \quad (14)$$

$$F_{400\_c} = \frac{1}{8}\left[\frac{4\sin^2\alpha_c}{r_c^2}\left(\frac{\cos^2\alpha_c}{r_c}-\frac{\cos\alpha_c}{R_c}\right)-\frac{1}{r_c}\left(\frac{\cos^2\alpha_c}{r_c}-\frac{\cos\alpha_c}{R_c}\right)^2+\frac{1}{R_c^2}\left(\frac{1}{r_c}-\frac{\cos\alpha_c}{R_c}\right)\right.$$
$$\left.+\frac{4\sin^2\alpha_c}{r_c'^2}\left(\frac{\cos^2\alpha_c}{r_c'}-\frac{\cos\alpha_c}{R_c}\right)-\frac{1}{r_c'}\left(\frac{\cos^2\alpha_c}{r_c'}-\frac{\cos\alpha_c}{R_c}\right)^2+\frac{1}{R_c^2}\left(\frac{1}{r_c'}-\frac{\cos\alpha_c}{R_c}\right)\right] \quad (15)$$

Where the reflection angle from the convex mirror is equal to the incident angle $\alpha_c$, $r_c$ and $r_c'$ are the object and image distance of the convex mirror respectively, $R_c$ is its meridional radius.

By setting $F_{200\_c} = 0$, it leads to:

$$R_c = \frac{2}{\cos\alpha(\frac{1}{r_c} + \frac{1}{r_c'})} < 0 \quad (16)$$

Since $r_c = L - d$ and $r_c' = (L-d)M_c$ ($-1 < M_c < 0$), where the convex mirror forms a reduced virtual image. So the overall F terms for the system consisting of a pre-convex mirror and a concave VLS grating could be recalculated by,

$$F_{200\_sum} = F_{200} \quad (17)$$

$$F_{300\_sum} = F_{300} + (-1)F_{300\_c} \times M_g \quad (18)$$

$$F_{400\_sum} = F_{400} + (-1)F_{400\_c} \times M_g \quad (19)$$

Where $M_g$ is the magnification of the grating (refer to Eq.(5)), since F term is proportional to the line width of the spectrum; the (-1) term in the formula is due to the virtual image created by the convex mirror (while it represents the real source of the grating effectively).

When the beam passes through the optical system, the optical aberration would broaden the beam size from the ideal spectral imaging distribution, the aberration broadening effect in the detector domain could be expressed as:

$$\Delta y_{ijk} = \frac{r'}{\cos\beta} \frac{\partial}{\partial w}\left[F_{ijk} w^i l^j\right] \quad (20)$$

Where $w$ is the illuminated meridional length of the grating, $l$ is the illuminated sagittal length, and $F_{ijk}$ defines the optical aberrations in various orders e.g. in Eq. (17-19)(the subscript $i$ or $j$ denotes the meridional or sagittal coordinate respectively, $k$ represents the orthogonal coordinate with $i$ and $j$).

Therefore, the meridional radius $R$ and coefficient $D_1$ of the VLS grating could be re-optimized by letting $r = d - r_c' > 0$ in Eq. (12) to obtain the best flat field for the whole spectral range, while $D_2$ and $D_3$ should be modified as well by solving Eq. (18-19) at the center wavelength $\lambda_0$.

Upon the above discussion, most of the parameters in the optical system could be determined, while among them $d$ and $M_c$ are special variables. In the next section, we will introduce a scheme to explore the desirable values of $d$ and $M_c$ to optimize the system design.

## 3. System optimization

### 3.1 The spot diagram and spectral distribution quality

In a system with pre-focusing (diverging) mirror and VLS grating, the optical aberration distribution is more complicated and difficult to calculate precisely. Even implementing the VLS grating, the perfect aberration compensation is hard to achieve, so the residual aberration terms would spread the spectral linewidth to reduce the resolving power of the system.

According to the discussion in previous sections (refer to $A_4$ in Fig.2), we find out:

a) The resolving power decreases with $|M_c|$ (magnification of the pre-convex mirror) monotonously for all spacing values of $d$, while too small $|M_c|$ should be eliminated in the design since it would induce too much large aberration to compensate.

b) The system prefers a larger d to deliver a relatively higher resolving powers. While the larger $d$ is, the further the pre-convex mirror is separated from the grating, leading to a larger illuminated area on it, which means advance grating manufacturing technique is in demand to enhance the effective optical area with considerably small fabrication error.

Keeping these in mind, the resolution enhanced spectrometer via implementing a pre-divergent mirror could be developed, and the system optimization should at least minimize optical aberrations to maintain a decent spectral imaging distribution. In order to evaluate the spectral distribution of the system for different parameter sets, we refer to the ray-tracing program and analyze the spot diagram on the detector plane. And the ratio of standard deviation of the meridional coordinates ($y_i$) of the outgoing rays and the line width of the diffraction beam distributed at the detector is used to calibrate the imaging quality at each specific wavelength:

$$Q = \frac{\sqrt{\frac{1}{N}\sum_{i=1}^{N}(y_i - \bar{y})^2}}{\sigma_d^{[FWHM]}} \quad (21)$$

Where $\bar{y}$ is the average value of $y_i$, $N$ is the total number of diffraction rays in simulation (here it is set to 10000), the denominator of Eq. (21) represents the ideal line width of the beam foot-print on the detector, and could be calculated by [24]:

$$\sigma_d^{[FWHM]} = \sigma_S^{[FWHM]} \frac{\cos\alpha}{\cos\beta} \frac{r'(\lambda)}{r} \frac{m}{\cos\theta} \quad (22)$$

Where $\theta$ is the defined as the angle in-between the central diffraction beam and the normal of the X-ray detector, $r$ and $r'(\lambda)$ are the object and image distances of the grating respectively.

Generally, the larger the value of $Q$, the greater the optical dispersion and the worse the imaging quality; and vice versa. The spot diagrams at 5nm for three different sets of $d$ and $M_c$ were obtained from Shadow ray-tracing program[26] and presented in Fig.3 for comparison, where the $Q$ value for each case were calculated to evaluate the corresponding spectral imaging quality.

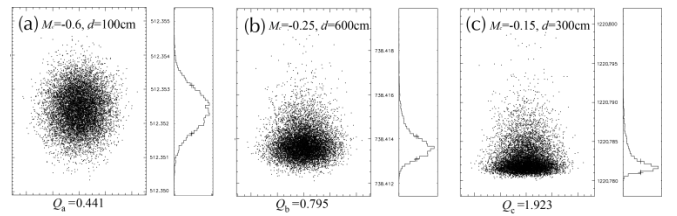

Fig.3. Comparison of the spectral distributions of the system at the detector domain, with 3 different sets of $d$ and $M_c$, where the wavelength is 5nm. The imaging quality is evaluated by the justified standard deviation of the meridional coordinates $Q$ of the outgoing rays (vertical distribution). The larger the $Q$ value, the worse the imaging quality; and vice versa. (a) $M_c$=-0.6, $d$=100cm, $Q_a$=0.441; (b) $M_c$=-0.25, $d$=600cm, $Q_b$=0.795; (c) $M_c$=-0.15, $d$=300cm, $Q_c \approx$1.923.

As depicted in Fig.3, the imaging quality of (a) is pretty good, exhibiting an evenly distributed and symmetric feature, while image quality of (b) and (c) are a lot worse; where the distribution of outgoing beam deviates from an ideal Gaussian peak, appearing certain degrees of asymmetry. The $Q$ value of the latter two ($Q_b$=0.795 or $Q_c$≈1.923) are much larger than the first one ($Q_a$=0.441), which implicates the system is not always optimized, especially when the aberrations in the system are not well corrected. Generally, the actual resolving power is significantly less than the ideal case, so we establish the criteria $Q$ to identify the realistic spectral quality for various cases. However, the parameters of $M_c$ and $d$ are dependant on each other, searching for an optimal set of parameters is not straightforward. So a machine learning scheme is introduced to narrow down the pool for exploring the various variables in demand and to improve the efficiency for identification of the optimal system, which will be discussed in next.

### 3.2 The system optimization through machine learning scheme

Following up the previous section, the machine learning scheme is organized as follow: $d$ and $M_c$ are set as the input variables, the rest of parameters of the optical system are either fixed or determined according to the input variables associatively, while the imaging quality $Q$ is the output. Through iterative modelling and learning, the machine could nicely predict the imaging quality of the system with different sets of parameters, thus approach to the best values of $d$ and $M_c$.

More specifically, the Support Vector Machine (SVM) is introduced to do the job, through implementing the structural risk minimization inductive principle to obtain generalization from a limited number of learning patterns to predict further more results [27, 28]. SVM has two main categories: Support Vector Classification (SVC) and Support Vector Regression (SVR)[29], here the latter is utilized to minimize the system errors to achieve generalized performance, where the computation is based on a linear regression function in a multi-dimensional space (>>3) while the input data are mapped via a nonlinear scheme. In current research, we adopted a powerful software LIBSVM and model developed by Chih-Chung Chang and Chih-Jen Lin [30].

Again, the parameters at the beginning of Chapter 2 were used: wavelength range 2~5nm, size of light source 50μm(r.m.s), beam divergence angle 20μrad (r.m.s), Gaussian type, $D_0$=24000ln/cm, both the incident angle of grating and convex mirror are set to 89°, $L$=30m and etc. Then various sets of $M_c$ and $d$ were used as the two input variables of the support vector machine for training. The rest of the parameters of the spectrometer could be calculated associatively, the spectral distribution and image quality were evaluated by the ray-tracing spot-diagram and the justified standard deviation $Q$. There are total 233 sets of $[M_c^{[i]}, d_i] \sim Q_i$ samples generated within certain restriction, where $i$ is the index of samples; among them, the first 200 samples selected randomly were input to LIBSVM for training and calibration, and the last 33 were used as verification. For a system with only two featured input variables, LIBSVM can easily gain convergence. More explicitly, the learning formula is reconstructed from 200 sets of samples (refer to Eq. (22), where only the first few terms are shown explicitly, $M_{c\_n}$, $d_n$ and $Q_n$ are normalized form of $M_c$, $d$ and $Q$ respectively).

$$Q_n(d_n, M_{c\_n}) = 208.0191\ e^{-1.4142(M_{c\_n}+0.5)^2 - 1.4142 d_n^2}$$
$$-86.4091\ e^{-1.4142(M_{c\_n}+1)^2 - 1.4142 d_n^2} \quad (22)$$
$$+360.2653\ e^{-1.4142(M_{c\_n}-1)^2 - 1.4142(d_n+1)^2}$$
$$-10.3757\ e^{-1.4142(M_{c\_n}+2.22e\text{-}16)^2 - 1.4142(d_n-1)^2}\ \ldots\ldots$$

Using this equation, various input values of $M_c$ and $d$, would lead to different $Q$ to predict the spectral image quality of the specific system. Thus the optimal set possessing the highest ideal resolving power while satisfying the $Q$ constraint could be identified. The general restrictions for the system optimization could be described in below:

$$\text{Max}\ A_4(d, M_c) = \frac{\lambda D_0}{\sigma_s^{FWHM} \cos\alpha}(L - d(1 + \frac{1}{M_c}))$$

$$\text{Subjected to}: \begin{cases} Q(d, M_c) < 0.51 \\ 100 < d < 2500 \\ -0.7 < M_c < -0.1 \end{cases}$$

Using the simple grid searching scheme, the best set of parameters were found: $M_c$=-0.427, $d$=1402cm. The optimization process is demonstrated in Fig.4. The blue mesh in Fig.4(a) shows $Q$ distribution profile with dependence on $d$ and $M_c$, and the regime for $Q(d, M_c) < 0.51$ (empirical value) meets the restriction for system optimization. By projecting it onto the plane $Q$=0, the effective domain to choose the valid $d$ and $M_c$ is determined. When $|M_c|$ is small ($|M_c|$<0.3), the optical elements spacing $d$ also need to be small to meet the constraint. On the other hand, the choices of 'd' are more flexible, when $|M_C|$ is relatively larger. Then the distribution profile of $A_4(d, M_c)$ are plotted in Fig.4(b), there is a trend of higher resolving power for smaller $|M_c|$ and bigger $d$. The colored curves in the plane of $A_4$=0 are associated with the equal-resolution contour from the $A_4$ profile, i.e. indicating all the available sets of $d$ and $M_C$ with identical ideal resolving power. Meanwhile the valid domain obtained from Fig.4(a) is plotted on the plane of $A_4$=0 against various contour lines of $A_4$. And it is not difficult to find out that the optimization approaches the contour line with a resolving power of 285000, which intersects with the effective domain to identify an optimal parameter set of $M_c$=-0.427, $d$=1402cm. The other parameters of the system were determined associatively and listed in Table.1.

It should be pointed out that the results above were obtained by machine learning the quality of spectral distribution ($Q$ function) at 5nm. Similarly, the machine learning scheme could be applied to the other wavelengths in the spectral range. Fig.4(c) demonstrates Q distribution with different sets of $M_c$ and $d$ at wavelengths of 2nm, 3.5nm and 5nm, only in the specific portion of the domain, 0.41<Q<0.55. It can be seen that within the effective domain (for system optimization), $Q_{2nm}$ (black stars) and $Q_{3.5nm}$ (red circles, central wavelength) have similar distribution profiles, while $Q_{5nm}$ (blue squares) are slightly larger than them. This indicates optimization of $Q_{5nm}$ is not just achieving an optimal system at the single wavelength of 5nm, the process would lead to an optimal system spanning for the entire "water window", i.e. 2-5nm.

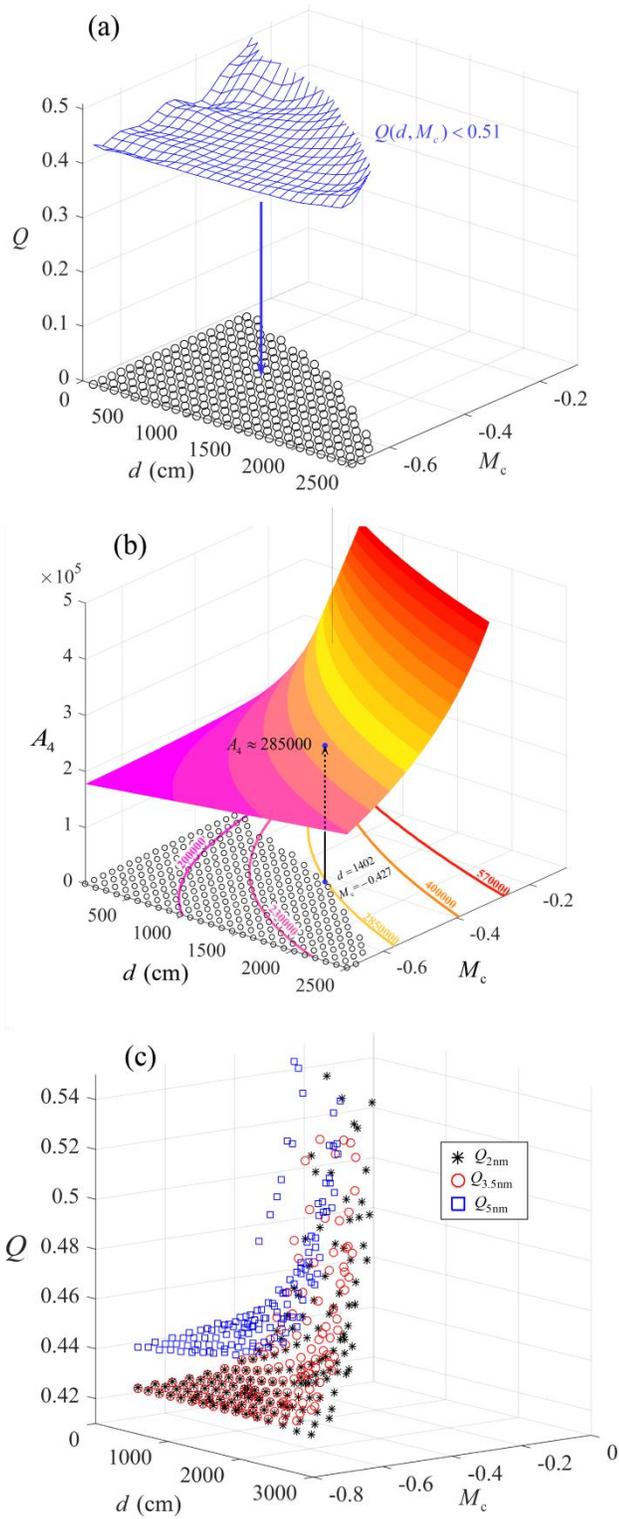

$Q(d, M_c) < 0.51$ gives the projection domain in the plane of $Q$=0, where the small circles on the plane indicate that only within this regime, the quality of the spectral distribution is good enough. (b) illustrates $A_4$ distribution profile, and the dependence with $d$ and $M_c$. The curves with various colours in the plane of $A_4$=0 are associated with the equal-resolution contour from the $A_4$ profile. Thus one set of the optimal parameters were found: $M_c$=-0.427, $d$=1402cm, corresponding to an ideal resolving power of 285000. The scatter plots in (c) shows Q distributions (0.41<Q<0.55) for the systems with different sets of $M_c$ and $d$ at various wavelengths: 2nm (stars), 3.5nm (circles) and 5nm (squares). Confining $Q_{5nm}$ below certain value ($Q_{5nm}$<0.51) means that the imaging quality of the entire spectral range is satisfied.

## 4. More comments on ray-tracing、aberration and fabrication errors

In previous section, we formulated a novel scheme for design of a resolution enhanced spectrometer, by implementing a pre-convex mirror to generate a reduced virtual image, which acts as an effectively real source for the VLS grating downstream. The aberrations of the convex mirror should also be considered and combined with the grating in system design and optimization. The SVM is used to explore the optimal parameters more efficiently, to eliminate the system's primary aberrations throughout the wavelength range to achieve extremely high resolving power with excellent spectral distribution simultaneously.

In order to evaluate the actual resolving power of a realistic spectrometer $A_4$, number of primary factors need to be considered and analysed. First of all, the spectral line width at the detector due to the light source size is (i.e. the ideal line width)[24]:

$$\Delta\lambda_s = \frac{\sigma_s^{(FWHM)} |M_c| \cos\alpha}{D_0 \cdot r} \quad (23)$$

Thus the ideal spectral resolution could be calculated by $A_{ideal} = \lambda / \Delta\lambda_S$, assuming a Gaussian beam in an aberration-free optical system, whose resolving power is mainly limited by the light source size $\sigma_s^{(FWHM)}$, enhanced by a factor of $1/|M_c|$ from $A_1$. While in a realistic optical system, the optical aberrations are non-negligible, which would broaden the spectral width distribution of an ideal Gaussian beam substantially, according to:

$$\Delta\lambda_{ijk} = \frac{\Delta y_{ijk} \cos\beta}{D_0 mr'} = \frac{\cos\beta}{D_0 mr'} \frac{r'}{\cos\beta} \frac{\partial}{\partial w}\left[F_{ijk} w^i l^j\right] = \frac{1}{mD_0} \frac{\partial}{\partial w}\left[F_{ijk} w^i l^j\right] \quad (24)$$

Where $\Delta y_{ijk}$ is the meridional beam size at the detector (refer to Eq. (20)), and the first few dominant aberration terms are (only for the meridional components, thus the sagittal index $l$=0):

$$\Delta\lambda_{200} = \frac{1}{D_0 m} 2w F_{200\_sum} \quad (25)$$

$$\Delta\lambda_{300} = \frac{1}{D_0 m} 3w^2 F_{300\_sum} \quad (26)$$

Fig4. Q-value represents the justified standard deviation of the meridional coordinates of the ray-tracing spot-diagram, and the resolving power of $A_4$ is given by Eq. (4); both of them are the crucial parameters for the system optimization. (a) shows $Q$ distribution profile (blue mesh), with dependence on $d$ and $M_c$. The restriction

$$\Delta\lambda_{400} = \frac{1}{D_0 m} 4w^3 F_{400\_sum} \quad (27)$$

The explicit expressions of $F_{200\_sum}$, $F_{300\_sum}$ and $F_{400\_sum}$ were already given in equations (17-19), which are independent of either $w$ or $l$.

For an optical system aiming for exceptional high spectral resolution, the requirements for the fabrication error (or height error) are very critical, including the slope error and surface roughness etc. for both the convex mirror and the grating, which broadens the spectral line width by[20],

$$\Delta\lambda_{CM}^{[SE]} = 2.355 \cdot SE_{CM} \frac{1}{D_0 m}(cos\alpha + cos\alpha) \quad (28)$$

$$\Delta\lambda_{G}^{[SE]} = 2.355 \cdot SE_{G} \frac{1}{D_0 m}(cos\alpha + cos\beta) \quad (29)$$

Where $SE_{CM}$ and $SE_G$ represent the meridional slope error of the convex mirror and grating, respectively. Assuming that they have an identical value i.e. $SE_{CM} = SE_G$, then the accumulative height error of the system is:

$$\Delta\lambda_{SE} = \Delta\lambda_{CM}^{[SE]} + \Delta\lambda_{G}^{[SE]} \quad (30)$$

The upper-bound of the spectral width due to the slope error (refer to Eq.(28-30)) could be set to that of the source size (refer to Eq.(23)), then the slope error of the optical element should be:

$$SE \leq \frac{|M_c|\sigma_s^{(FWHM)}\cos\alpha}{2.355 \cdot r \cdot (3cos\alpha + cos\beta)} \quad (31)$$

Using the source size and diffraction angle $\beta$ at 5nm, the expected slope error should be smaller than 0.086μrad. Currently the fabrication requirement for $SE \leq 0.1$ μrad is very challenging and rare, even for the most advanced grating manufacturing technique (there were reports about achieving an optical slope error of better than 0.05μrad though [18]). Since our ultimate goal is to develop a broadband spectrometer with exceptional resolution (>300000), it is worthwhile to demand the cutting-edge grating fabrication technology.

When all the effects in the realistic spectrometer are included, the resolution can be re-calculated:

$$A_{theory} = \frac{\lambda}{\Delta\lambda_{sum}} \approx \lambda[\Delta\lambda_S^2 + (\Delta\lambda_{200} + \Delta\lambda_{300} + \Delta\lambda_{400})^2 + \Delta\lambda_{SE}^2]^{-1/2} \quad (32)$$

And the spectrometer model in Table 1 could be used to calculate the various spectral distribution terms via implementing equations (23), (25-27) and (28-30), and the results are shown in Fig.5(a). The source size term $\Delta\lambda_S$ seems to be dominating, almost constant within the spectral range (since the source size is assumed to be constant throughout the spectral range). The slope-error term $\Delta\lambda_{SE}$ is the second largest component. The spectral broadenings due to three primary aberration components ($\Delta\lambda_{200}$, $\Delta\lambda_{300}$ or $\Delta\lambda_{400}$) are relatively small and well confined.

Table.1 Design parameters of the optimized spectrometer ($A_4$)

| Source | Type | Size | Divergence angle | L |
|---|---|---|---|---|
|  | Gaussian | 50μm (rms) | 20μrad (rms) | 3000cm |
| Convex Mirror | $\alpha_c$ | $r_c$ | $r_c'$ | |
|  | 89° | 1598cm | -682cm | |
|  | d | $M_c$ | $R_c$ | |
|  | 1402cm | -0.427 | -136470cm | |
| Concave VLSG | α | r | r' | |
|  | 89° | 2084cm | 3100cm | |
|  | R | VLS coefficient | | |
|  | 270650 cm | $D_0$=24000ln/cm $D_1$=14.55ln/cm2 $D_2$=0.0065ln/cm3 $D_3$=3.689e-6 ln/cm4 | | |
| Footprint (FWHM) on the convex mirror surface | $w_{CM}$ | Footprint (FWHM) on the grating surface | $w_G$ | |
|  | 4.47cm |  | 13.43cm | |
| Required Slope Error (meridional) | $SE_{CM}$ | $SE_G$ | | |
|  | <0.086μrad | <0.086μrad | | |

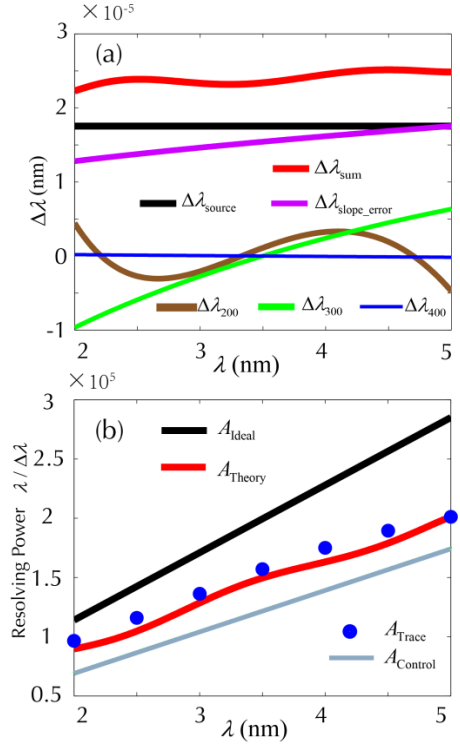

Fig.5 (a)The calculated results of the major factors which influence the resolving power of the spectrometer, including the source size (black), the optical fabrication error (purple), the optical aberrations – defocus (brown), coma (green), spherical aberration (blue) and the overall (thick red). The corresponding resolving powers of (a) are calculated and presented in (b), where three different types of the spectral resolutions are: $A_{ideal} = \lambda/\Delta\lambda_{so}$ (black), $A_{theory} = \lambda/\Delta\lambda_{sum}$ (red), $A_{trace}$ (dot-blue) obtained from the ray-tracing program; and a control signal $A_{control}$ (grey) is plot in the same spectral range, calculated by Eq. (1) i.e. ideal resolution of $A_1$ with identical $L$.

The corresponding resolving powers for various terms in Figs. 5(a) are exhibited in Figs. 5(b), where the ideal spectral resolution $A_{ideal} = \lambda/\Delta\lambda_S$ (thick black), the theoretical resolution (thick red), $A_{theory} = \lambda/\Delta\lambda_{sum}$, the result from the ray-tracing program $A_{trace}$ (discrete blue disks) and a control group $A_{control}$ (grey) calculated by Eq.(1) using an identical $L$, are overlaid for comparison. Obviously, the theoretical resolving power of a realistic spectrometer $A_4$ (thick red) including the contribution from slope error and optical aberrations, is still considerably larger than the ideal resolving power of a single grating spectrometer $A_1$ (grey). This indicates that if the precision of grating manufacturing were pushed to the extreme limit, the system would achieve even higher spectral resolution, approaching to the ideal value $A_{ideal}$.

Additionally, the ray-tracing results for the spectrometer with configuration in Table.1 are presented in Fig. 6. The bottom part of the figure shows the spectral distributions at the optimal detector plane throughout "water-window" range (i.e. 2-5 nm), where the length scales in the meridional (2000mm) and sagittal (20mm) directions are quite different. While Fig.6 (a-d) exhibit the spectral distribution and resolution at each individual wavelength respectively (2, 3, 4, 5 nm in terms of $\lambda$ and $\lambda+\Delta\lambda$), each in an identical detector domain of rectangle: 20mm(sagittal) × 0.1mm(meridional). Especially the FWHM beam widths for each wavelength in meridional coordinate are illustrated in specific sub-plots, which are set to be larger than the typical pixel size of a CCD detector ~10μm, to guarantee the realization of the spectral resolution. According to Eq.(6), the image distance of the grating $r'$ should be at least about 30 meters for an optimal spectrometer $A_4$ to achieve the ideal resolving power of 300,000. This means that the length scale of the outgoing beam of the spectrometer would be very large, so as the detector range. While our design delivers an excellent flat field crossing throughout the spectral range, the CCD detector could be mounted and scanned on a pretty much straight guide-rail to cover the entire spectrum.

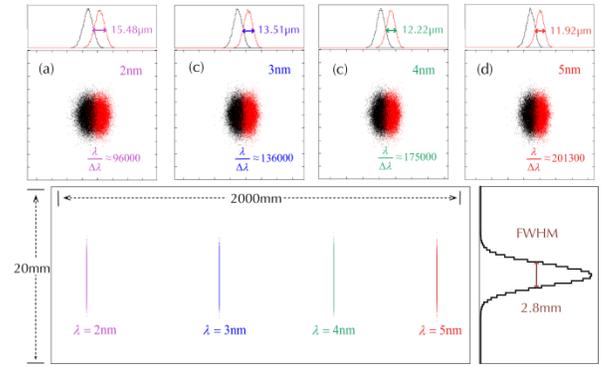

Fig.6. The ray-tracing results for the spectrometer configuration in Table.1. The spectral profile distributions at the optimal detector plane for the full wavelength range (2-5 nm) are demonstrated at the lower part of the figure, where the detector needs to scan to cover a area of 2000mm (meridional)×20 mm (sagittal). The ray-tracing results for each wavelength of 2 nm, 3 nm, 4 nm and 5 nm are presented in (a) to (d), spanning an identical 'detector domain' of 20mm(vertical)×0.1mm(horizontal) for each: the meridional size of the spectrograph is 11-16 "μm" (FWHM), the sagittal size is about 2.8mm (FWHM). Then the resolution power at various wavelengths are presented: (a) 96000 at 2 nm, (b) 136000 at 3 nm, (c) 175000 at 4 nm, and (d) 201300 at 5 nm.

## 5. Discussion and Conclusion

In summary, we report a novel spectrometer design in combination with a cylindrically convex pre-mirror and a cylindrically concave VLS grating (both in meridional). The design could not only provide a decent flat field at the detector domain, but also enhance the resolving power substantially. Our main findings in the current research are: 1) If a convex mirror is inserted in-between the light source and the grating to create a reduced virtual image (acting as a secondary real source point for the grating), the resolution of the system would be enhanced. 2) Generally, if a pre-mirror (convex or concave) is inserted upstream the grating, its optical aberration should be included and justified (e.g. the magnification, creating a real or virtual image), in order to calculate and compensate the overall aberration of the system accurately. 3) A realistic optical system always possesses errors e.g. optical aberrations and fabrication errors, thus the beam spectral distribution would be broader than and deviate from an aberration free ideal Gaussian distribution; and the standard deviation of the outgoing beam's spot diagram could be used to reflect the image quality. 4) The support vector machines can quickly learn from the input data and reconstruct the prediction formula to explore the optimal system with excellent imaging quality. Implement the nonlinear programming script, the optimized parameter set of $M_c$ and $d$, associated with the highest resolving power could be identified. 5) For a spectrometer system with extremely high resolving power, it always has very high demands for the precise optical manufacturing, i.e. requiring exceptionally small slope error or surface roughness for the optical elements in the system.

The position and magnification of the pre-convex mirror are the crucial parameters in the current spectrometer design, which also constrain the selection for the object and image distances of the grating, thus reduce the number of variables for system optimization. Then implementation of the machine learning scheme could explore and identify the optimal system delivering the excellent resolution while maintaining the minimal optical aberrations, with pretty high efficiency. Although we mainly discussed the spectrometer design in "water window", the algorithm owns universal adaptability, which could be easily extended in much broader photon-energy (or wavelength) range through an appropriate modification to the design parameters. And it is also feasible to utilize the scheme to develop a high-performance grating monochromator simply by putting a fine slit right across the focal plane of the diffraction beam. And, the scheme could be straightforward used on many types of experiments which pursue absolutely high spectral resolution. Most remarkably, the outstanding advantage of the current spectrometer design is, the high resolving power could be realized at rather large source size, no need to tight focus the beam or apply confinement slit!

## FUNDING SOURCES AND ACKNOWLEDGMENTS

**Funding.** National Science Foundation of China (NSFC) (11475249), and Youth 1000-Talent Program in China (Y326021061).

**Acknowledgment**. The authors thank for the staff and facility support from the Department of Free Electron Laser Science & Technology, Shanghai Institute of Applied Physics, Chinese Academy of Sciences.

See Supplement 1 for supporting content.

## REFERENCES

1. L. J. P. Ament, M. van Veenendaal, T. P. Devereaux, J. P. Hill, and J. van den Brink, "Resonant inelastic x-ray scattering studies of elementary excitations," Reviews of Modern Physics 83, 705-767 (2011).
2. P. Kuiper, J.-H. Guo, C. Såthe, L.-C. Duda, J. Nordgren, J. Pothuizen, F. De Groot, and G. A. Sawatzky, "Resonant X-ray raman spectra of Cu dd excitations in Sr 2 CuO 2 Cl 2," Physical review letters 80, 5204 (1998).
3. Y. Harada, T. Kinugasa, R. Eguchi, M. Matsubara, A. Kotani, M. Watanabe, A. Yagishita, and S. Shin, "Polarization dependence of soft-x-ray Raman scattering at the L edge of TiO 2," Physical Review B 61, 12854 (2000).
4. L. Braicovich, J. Van den Brink, V. Bisogni, M. M. Sala, L. Ament, N. Brookes, G. De Luca, M. Salluzzo, T. Schmitt, and V. Strocov, "Magnetic excitations and phase separation in the underdoped La 2− x Sr x CuO 4 superconductor measured by resonant inelastic X-ray scattering," Physical review letters 104, 077002 (2010).
5. M. Guarise, B. Dalla Piazza, M. M. Sala, G. Ghiringhelli, L. Braicovich, H. Berger, J. N. Hancock, D. Van Der Marel, T. Schmitt, and V. Strocov, "Measurement of magnetic excitations in the two-dimensional antiferromagnetic Sr 2 CuO 2 Cl 2 insulator using resonant X-ray scattering: Evidence for extended interactions," Physical review letters 105, 157006 (2010).
6. K.-J. Zhou, Y.-B. Huang, C. Monney, X. Dai, V. N. Strocov, N.-L. Wang, Z.-G. Chen, C. Zhang, P. Dai, and L. Patthey, "Persistent high-energy spin excitations in iron-pnictide superconductors," Nature communications 4, 1470 (2013).
7. F. Hennies, A. Pietzsch, M. Berglund, A. Föhlisch, T. Schmitt, V. Strocov, H. O. Karlsson, J. Andersson, and J.-E. Rubensson, "Resonant inelastic scattering spectra of free molecules with vibrational resolution," Physical review letters 104, 193002 (2010).
8. A. Pietzsch, Y.-P. Sun, F. Hennies, Z. Rinkevicius, H. O. Karlsson, T. Schmitt, V. N. Strocov, J. Andersson, B. Kennedy, and J. Schlappa, "Spatial quantum beats in vibrational resonant inelastic soft x-ray scattering at dissociating states in oxygen," Physical review letters 106, 153004 (2011).
9. T. Namioka, "Theory of the concave grating. I," Josa 49, 446-460 (1959).
10. P.-z. Fan, Z.-q. Zhang, J.-z. Zhou, R.-s. Jin, Z.-z. Xu, and X. Guo, "Stigmatic grazing-incidence flat-field grating spectrograph," Applied optics 31, 6720-6723 (1992).
11. G. Xiong, Z. Hu, H. Li, Y. Zhao, W. Shang, T. Zhu, M. Wei, G. Yang, J. Zhang, and J. Yang, "One-dimensional space resolving flat-field holographic grating soft x-ray framing camera spectrograph for laser plasma diagnostics," Review of Scientific Instruments 82, 043109 (2011).
12. T. Harada and T. Kita, "Mechanically ruled aberration-corrected concave gratings," Applied Optics 19, 3987-3993 (1980).
13. N. Nakano, H. Kuroda, T. Kita, and T. Harada, "Development of a flat-field grazing-incidence XUV spectrometer and its application in picosecond XUV spectroscopy," Applied optics 23, 2386-2392 (1984).
14. I. W. Choi, J. U. Lee, and C. H. Nam, "Space-resolving flat-field extreme ultraviolet spectrograph system and its aberration analysis with wave-front aberration," Applied optics 36, 1457-1466 (1997).
15. M. C. Hettrick, J. H. Underwood, P. J. Batson, and M. J. Eckart, "Resolving power of 35,000 (5 mÅ) in the extreme ultraviolet employing a grazing incidence spectrometer," Appl Opt 27, 200-202 (1988).
16. C. Hague, J. Underwood, A. Avila, R. Delaunay, H. Ringuenet, M. Marsi, and M. Sacchi, "Plane-grating flat-field soft x-ray spectrometer," Review of scientific instruments 76, 023110 (2005).
17. G. Tondello, "The use of a toroidal mirror as a focusing element for a stigmatic grazing incidence spectrometer," Journal of Modern Optics 26, 357-371 (1979).
18. J. Dvorak, I. Jarrige, V. Bisogni, S. Coburn, and W. Leonhardt, "Towards 10 meV resolution: The design of an ultrahigh resolution soft X-ray RIXS spectrometer," Review of Scientific Instruments 87, 115109 (2016).
19. T. Warwick, Y. D. Chuang, D. L. Voronov, and H. A. Padmore, "A multiplexed high-resolution imaging spectrometer for resonant inelastic soft X-ray scattering spectroscopy," J Synchrotron Radiat 21, 736-743 (2014).
20. Y. D. Chuang, Y. C. Shao, A. Cruz, K. Hanzel, A. Brown, A. Frano, R. Qiao, B. Smith, E. Domning, S. W. Huang, L. A. Wray, W. S. Lee, Z. X. Shen, T. P. Devereaux, J. W. Chiou, W. F. Pong, V. V. Yashchuk, E. Gullikson, R. Reininger, W. Yang, J. Guo, R. Duarte, and Z. Hussain, "Modular soft x-ray spectrometer for applications in energy sciences and quantum materials," The Review of scientific instruments 88, 013110 (2017).
21. H. Wolter, "Spiegelsysteme streifenden Einfalls als abbildende Optiken für Röntgenstrahlen," Annalen der Physik 445, 94-114 (1952).
22. T. T. Saha, "Aberrations for grazing incidence telescopes," Applied optics 27, 1492-1498 (1988).
23. T. T. Saha, "Transverse ray aberrations for paraboloid–hyperboloid telescopes," Applied optics 24, 1856-1863 (1985).
24. Z. Li and B. Li, "A sagittally confined high-resolution spectrometer in the water window'," Journal of synchrotron radiation 25, 738-747 (2018).

25. J. A. Samson, D. L. Ederer, T. Lucatorto, and M. De Graef, Vacuum ultraviolet spectroscopy I (Academic Press, 1998), Vol. 31.
26. M. Sanchez del Rio, N. Canestrari, F. Jiang, and F. Cerrina, "SHADOW3: a new version of the synchrotron X-ray optics modelling package," Journal of synchrotron radiation 18, 708-716 (2011).
27. V. Vapnik, "Pattern recognition using generalized portrait method," Automation and remote control 24, 774-780 (1963).
28. V. Vapnik, "A note one class of perceptrons," Automation and remote control (1964).
29. V. Vapnik, "Support vector method for function approximation," Advances in neural information processing systems 9, 281–287 (2001).
30. C.-C. Chang and C.-J. Lin, "LIBSVM: a library for support vector machines," ACM transactions on intelligent systems and technology (TIST) 2, 27 (2011).